# Towards Economic Models for MOOC Pricing Strategy Design

Yongzheng Jia[1], Zhengyang Song[1], Xiaolan Bai[2] and Wei Xu[1]

[1]Institute of Interdisciplinary Information Sciences, Tsinghua University
[2]Faculty of Education, the University of Hong Kong

**Abstract.** MOOCs have brought unprecedented opportunities of making high-quality courses accessible to everybody. However, from the business point of view, MOOCs are often challenged for lacking of sustainable business models, and academic research for marketing strategies of MOOCs is also a blind spot currently. In this work, we try to formulate the business models and pricing strategies in a structured and scientific way. Based on both theoretical research and real marketing data analysis from a MOOC platform, we present the insights of the pricing strategies for existing MOOC markets. We focus on the pricing strategies for verified certificates in the B2C markets, and also give ideas of modeling the course sub-licensing services in B2B markets.

## 1 Introduction

Going on eight years since *massive open online courses* (MOOCs) first entered the scene, MOOCs go from the cameras at the back of college classrooms to new forms of online education ecosystems in the global industry. MOOCs bring a revolution to the worldwide higher education for growing opportunities in the verticals such as online education, lifelong learning, professional training, by offering freely accessible college education to everyone. However, MOOCs have also been criticized heavily by academics and industries for operational sustainability, low completion rate, unprofessional teaching methods, as well as the corporatization of higher education.

MOOC is an ecosystem involving efforts from many parties. The *MOOC platforms* are the core of the ecosystem. Every MOOC platform is a market place where *MOOC producers* (usually universities) delivers their MOOCs to the *users*. The users in the MOOC ecosystem consist of both *Internet users* and *institutional users*. The platforms also offer the users with various value-added *MOOC services* to increase their profitability. There are both for-profit platforms (e.g. Coursera and Udacitity) and non-profit platforms (e.g. edX and FUN), and the format of the *MOOC services* varies from platform to platform.

One critical issue to discuss is how does the MOOC ecosystem stay financially stable. In the beginning, both public and private sector fundings flood into these MOOC platforms so they can focus on adding contents and expanding the market share. However, after a few years, people not only want MOOC platforms to be financially independent but also provide financial incentives to the MOOC providers. Some MOOC platforms (e.g. xuetangX) are still in the primary stage of marketing, and struggling to make money and show their investors that they can be sustainable, or at least reach the breakeven point in the future.

Unfortunately, there is few academic research formally analyzing the business models for MOOCs in scientific and structured ways. This research is based on our research and the first-hand operation experience in industry. We focus on the pricing strategies of MOOC services in this paper, since the pricing strategy design is a key component of the MOOC business models, and also a common interest of the online education community, networking community, as well as researchers from marketing and economics.

There are two existing business models of MOOC platforms with totally distinct pricing strategy, the *business-to-customer* (B2C) model and *business-to-business* (B2B) model. Unlike the peer-education platforms (e.g. Skillshare and Udemy), there is usually no consumer-to-consumer model (C2C) as MOOCs are usually provided by organizations instead of individuals. While by definition, the MOOCs are free and open-to-all, the MOOC platforms sell value-added *MOOC services* for profit, and it is a common model in Internet services called the *freemium strategy*.

In the B2C markets, MOOC platforms make money by selling services to the *Internet users*, such as *verified certificates*, *specializations* and *online degrees*. In the B2B markets, MOOC services can be used as the form of *Small Private Online Courses* (i.e. SPOCs). The traditional in-classroom higher education can be transferred into a blended mode by using the high-quality MOOC content, and MOOC platforms can make money from *sub-licensing* the content to institutional users, and providing *bundled education services* including content customization, teaching assistant services, SaaS services, and technical supports.

In this work, we discuss our initial attempts to construct reasonable mathematical formulations for the pricing models based on our marketing experience. Due to the space reason, we focus on the pricing model for *verified certificates* in B2C markets. We summarize the key ideas of this paper as following:

*First,* We focus on the theoretical framework for the pricing strategies for the *verified certificates* in B2C markets. We first propose the basic model of certificate pricing, to solve the case of pricing the certificates for a single MOOC on one MOOC platform without competition. We analyze the profit maximization pricing strategies for MOOC platform and the market equilibrium with maximum social welfare. We further review the user-platform interaction of certificate pricing by using the *Stackelberg game*. Then we analyze the case of considering the purchasing power of users with budget constraints. In this model, we formulate a utility maximization framework to depict the consumers' buying behavior, and we can use the pricing strategies to increase per-user revenue. Finally, we summarize the model of bundled course services and their business initiatives.

*Second,* We analyze the sales data of certificates from 1236 real MOOCs based on our model. By using data-driven marketing approaches, we get some business and education insights of the users' behavior. We first present an overview of the marketing performance of selling certificates, and further analyze the users' behavior in different cases: the users' *willingness to pay* (i.e. WTP) for best-selling MOOCs and MOOCs with highest payment rate, as well as the users' behavior when a MOOC is repeatedly offered.

*Third,* We further present the future directions of the work on designing the pricing strategies. There are some other factors which may affect the B2C market in real settings and we should consider them in industry. We also propose the ideas of modeling the B2B sub-licensing markets, and the hybrid model of B2B2C (business-to-business-to-customer) with *cross-platform MOOC exchange and internationalization*. We will try to solve these problems in our future work.

The rest of the paper is organized as follows. We review related work in Section 2. We present the pricing strategy for B2C markets in Section 3 and analyze the real marketing data in Section 4. Finally, we present the directions of our future work in Section 5 and conclude the paper in Section 6.

## 2 Related Work

To the best of our knowledge, our work is the first to study the business model and marketing strategy of MOOCs with both theoretical models and data-driven analysis. There are many discussions on the business model and sustainability of MOOCs from the industry and the media. For instance, [1] shows the latest experience of finding niche and business model for MOOC in 2016, and [2] summarizes the business innovations and landscape changes for MOOC in 2016. Existing academic work on MOOC business model is based on case studies, surveys, and other social science methodologies. [3] presents the ideas of involving adaptive learning into the business model design of MOOCs. [4] shows the ideas of designing sustainable MOOC business models by carefully reviewing them for both US and Europe-based MOOC aggregators.

From a theoretical perspective, literatures in multiple fields give us ideas on analyzing pricing strategies by applying economics, optimization theory, and game theory. [5] presents a generalized *Smart Data Pricing* scheme of pricing the network applications to increase efficiency and cope with increased network congestion. [6] and [7] show the flat-rate pricing scheme for data traffic to obtain higher profits with game theory. To analyze the user behavior by economics, [8] reviews the methodologies of WTP estimation in marketing science, and [9] is a classical structural model of demand estimation in economics.

## 3 B2C Business Model

The fundamental B2C model of MOOC platforms is to make money from the Internet users with a *freemium* strategy: The basic materials of MOOCs are open and free to all users, and the MOOC platforms also offer fee-based *online value-added services* to the users. The basic strategy is to grow the user base of the platform, and then try to cultivate the users' payment habits with online marketing strategies. In this section, we focus on the pricing models of the *verified certificates*, which is the fundamental B2C value-added services. We mainly discuss two models of certificate pricing: One pricing strategy is to maximize the total profit from each MOOC, and the other is to increase the profit from each paying user.

### 3.1 The Basic Model of Certificate Pricing

We observe the basic market structure of the verified certificates by modeling the following straightforward scenario. We consider that a MOOC (labeled as $\mathcal{M}$) is released on only one MOOC platform. Furthermore, we ignore the competitive relationship across different MOOCs (e.g. MOOC $\mathcal{M}$ may be a machine learning MOOC on Coursera, and there is another similar machine learning MOOC on edX). To simplify the market structure, we treat the MOOC producer (i.e. content providers) of $\mathcal{M}$ and the MOOC platform as a single entity[1] (i.e. the seller in the market).

Therefore the market for $\mathcal{M}$ incorporates a single seller (i.e. the MOOC platforms) and a set of users (i.e. learners on the Internet). Suppose MOOC $\mathcal{M}$ offers the verified certificates with price $p$ for the paying users. We first consider users' decisions of whether to buy the verified certificate of MOOC $\mathcal{M}$. In modeling the user behavior, we suppose that each user acts so as to maximize her net benefit (i.e. consumer surplus), denoted by $U_j(x_j, p)$ for each customer $j \in \{1, 2, \cdots, J\}$ and $x_j \in \{0, 1\}$. The function $U_j$ is the net benefit to a consumer from the utility received in buying the verified certificate or just audit the course. We further use $\bar{V}_j$ to denote the utility to user $j$ of just auditing MOOC $\mathcal{M}$ or earning a free certificate, and use $V_j$ to denote the utility to user $j$ of taking the course and earning a verified certificate (i.e. the *willingness to pay* for the verified certificate). Then we formulate $U_j(x_j, p)$ as following:

$$U_j(0, p) = \bar{V}_j, \quad U_j(1, p) = V_j - p, \quad \forall j \in \{1, 2, \cdots, J\} \tag{1}$$

We then denote the decision of user $j$ under price $p$ as $x_j^*(p)$ to maximize her net benefit, which is also the *demand functions* for user $j$, and we can calculate $x^*(p)$ as follows:

$$x_j^*(p) = \begin{cases} 1 & \text{if } U_j(1, p) > U_j(0, p) \\ 0 & \text{otherwise} \end{cases} \quad \forall j \in \{1, 2, \cdots, J\} \tag{2}$$

By adding up the demand function of $x_j^*(p)$ for all the users $j \in \{1, 2, \cdots, J\}$, we obtain the *aggregate demand function* of $D(p) = \sum_{j=1}^{J} x_j^*(p)$ to capture the total demand of MOOC $\mathcal{M}$ in the market.

---

[1] The MOOC producers and the platforms build the collaboration based on agreements with revenue sharing terms. So in a single-seller market, we can treat the MOOC producer and the platform as a unity, without considering their internal interest exchanges.

Then we consider the strategy of the MOOC platform. We first observe the cost of the verified certificates. Due to the *economies of scales*[10] for Internet services, the MOOC services has huge fixed cost (including high production cost for a MOOC) but very low marginal cost. To offer verified certificate to one more paying user, the platform will only incur traffic cost, verification cost (i.e. identify the authentic user) and cost for some other value-added services. The marginal cost is small and fixed, we denote it as $\bar{c}$. Therefore the price $p$ should satisfy $p > \bar{c}$. If we can figure out the aggregate demand function of MOOC $\mathcal{M}$, then we can get a profit maximization pricing strategy from the following theorem:

**Theorem 1.** *For the basic model of verified certificate, the best (i.e. profit maximization) pricing strategy for MOOC $\mathcal{M}$ is:*

$$\bar{p} = argmax_p[D(p) \cdot (p - \bar{c})] \tag{3}$$

*Where $\bar{p}$ is the platform's best pricing strategy for MOOC $\mathcal{M}$.*

We then analyze the market equilibrium. As the marginal cost $\bar{c}$ is a constant, the market equilibrium occurs at $p = \bar{c}$. One often-analyzed property of this equilibrium is the *social welfare*, which is the sum of the producer surplus and consumer surplus in the market. We use $\mathcal{SW}(p)$ to denote the social welfare at price $p$, and we have:

$$\mathcal{SW}(p) = \sum_{j \in [J]} U_j(1, p) + \sum_{j \in [J]} x_j^*(p) \cdot (p - \bar{c}) \tag{4}$$

$\mathcal{SW}(p)$ will get its maximum value at the *market equilibrium price* when $p = \bar{c}$ in a perfectly competitive market. We can see that when the MOOC market is highly competitive, the net profit of the platform may diminish. However, MOOC market is at least an oligopoly market. There are limited number of MOOC producers and MOOC platforms in the market. For high-quality MOOC content, the number of competitive players is even smaller. Therefore in the MOOC market, users act as price-takers and have weak bargaining power against the platforms.

The above reasoning, in which a platform chooses a price to offer subject to users' behavior as a function of the price chosen, is an example of a *game* between users and platform. In such a game, several players interact with each other, and each player acts to maximize her utility, which may be influenced by other players decisions. For instance, in this basic model, we only consider the interaction between the users and MOOC platform for the pricing strategy. We can also consider the interaction between MOOC platforms. This idea leads us to think about some basic principles of game theory in relation to the pricing strategies of the verified certificates.

### 3.2 A Game-Theoretic View

To demonstrate some of the basics of game theory, we again consider the example above. The user-platform interaction is an example of a *Stackelberg game* [11] in which one player, the "leader" (i.e. platform), makes a decision (i.e., the platform sets a price $p$ for course $\mathcal{M}$'s verified certificate). The remaining players, or "followers" (i.e. users), then make their decisions based on the leaders actions. In the basic model of MOOC certificate pricing, it reflects that the price $p$ will influence the users' decision of whether to buy the certificate (i.e. the demand function $x_j^*(p)$). Stackelberg games often arise in user-platform interactions of the network economy, and we can use the method of *backwards induction* [12] to analyze the Stackelberg games: First, we computes the followers actions as a function of the leaders decision (in our example, we compute the function $x_j^*(p)$ for user $j \in [J]$). The followers' actions are also called the best response to the leader. Then in the second step, the leader (i.e. platform) takes these actions into account and makes its decision to best respond the users.

Due to the space reason, we omit the analysis of Stackelberg game in details. From the ideas of Stackelberg game, we can see that the best response to the followers of the platform depends on the distribution of the demand function $x_j^*(p)$, and furthermore, the WTP distribution of the verified certificates and the utility gained from auditing the course. In practice, we can also use the ideas from the *backwards induction* to develop experiments to estimate the WTP of the users: The platform can dynamically change the price for certificates (e.g. make a discount) to figure out the WTP distribution at each price level.

### 3.3 Taking Multiple Courses with Budget Constraints

From the above basic model for certificate pricing, we consider a more complicated model in which each user buys multiple course certificates with budget constraints. Consider a user $j \in [J]$, she plans to take a number of courses on one MOOC platform with a fixed budget constraint $B_j$ during a certain period (e.g. a month or a semester). Due to time limitation, she can take at most $K_j$ MOOCs. There is a total of $M$ courses in the market. Her WTP for the verified certificate of course $m \in [M]$ is $V_{j,m}$ and her utility of audit course $m$ is $\bar{V}_{j,m}$. The price for course $m$'s certificate is $p_m$. Similar to the basic model, we use $x_{j,m}(p_m)$ to denote whether user $j$ will pay for the certificate of course $m$ under price $p_m$. Then the user $j$'s behavior can be formulated by solving an integer programming as follows:

$$Z : \text{maximize} \sum_{m \in [M]} x_{j,m}(p_m) \cdot (V_{j,m} - p_m) \tag{5}$$

s.t.

$$x_{j,m}(p_m) \cdot (V_{j,m} - \bar{V}_{j,m} - p_m) \geq 0, \quad \forall j \in [J]; \tag{6a}$$

$$\sum_{m \in [M]} x_{j,m}(p_m) \cdot p_m \leq B_j, \quad \forall j \in [J]; \tag{6b}$$

$$\sum_{m \in [M]} x_{j,m}(p_m) \leq K_j, \quad \forall j \in [J]; \tag{6c}$$

$$x_{j,m}(p_m) \in \{0,1\}, \quad \forall m \in [M], j \in [J]. \tag{6d}$$

The objective function (5) is the total net benefit gained of user $j$. Constraint 6a guarantees that user $j$ is better-off from buying the certificate, and (6d) is the budget constraint. The user $j$'s strategy is a vector of $x_{j,m}(p_m)$ of whether to buy the certificate for course $m$ or not.

The user $j$'s strategy is the optimal solution of (5), then we try to calculate the best pricing strategy for the platform. Note that when $p_m$ is different for each course, the problem is hard to solve and we will omit the discussions here. However, when $p_m$ is the same for each course $m \in [M]$. We can estimate the demand function of user $j$ as follows:

**Theorem 2.** *If $p_m$ is same for each course (i.e. $p_m = p, \forall m \in [M]$), the platform tries to maximize the total number of certificates bought from user $j$. The demand function of user $j$ can be calculated by a function of $p$, $\{V_{j,m}\}_{m \in [M]}$, $\bar{V}_{j,m}$, $K_j$ and $B_j$, such that:*

$$D_j(p) = \sum_{m \in [M]} x_{j,m}(p_m) = \mathcal{F}_j\left(p, B_j, K_j, \{V_{j,m}\}_{m \in [M]}, \{\bar{V}_{j,m}\}_{m \in [M]}\right) \tag{7}$$

*and the aggregate demand function is $D(p) = \sum_{j \in [J]} D_j(p)$*

If we can estimate the WTP for each course of user $j$ and the user's budget distribution, we can use the methodology from Theorem 1 to design the platform's pricing strategy for profit maximization. From this model of analyzing user's purchasing behavior with budget constraints, we get the solution of maximizing the total profit from each paying user.

### 3.4 Bundled Course Services

Offering non-free verified certificate is the initial and basic business model for the B2C markets. For Coursera, the certificate services expand quickly of featuring a flat-rate price of $49 for all certificates. Revenue estimates suggest that Certificates generated between $8 and $12 million for Coursera in 2014 [13].

In practice, the MOOC platforms always put some of their courses together and add more value-added services to form the *bundled course services*, such as the *Specializations* on Coursera (or the *XSeries* on edX), the *Online Micro Masters* on edX (or the *Nano-degrees* on Udacity), the *Advanced Placement* (i.e. AP) courses and so forth. Offering the *bundled course services* is a strategy to improve the quality of value-added service and provide opportunities to charge the users with higher prices. It is a direction of our future work to model the bundled course services and show their performance.

## 4 Data-Driven Analysis

In this section, we present the results of applying our models to analyze the real sales data from a MOOC platform with millions of users and give some insights using data-driven marketing methods.

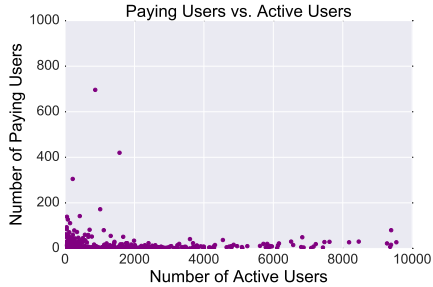 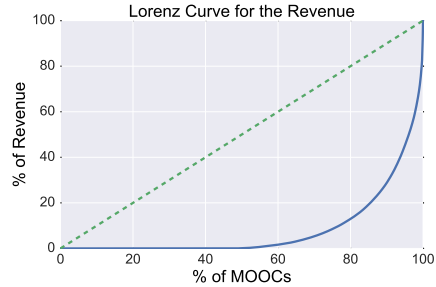

**Fig. 1.** Relationship between Paying Users and Active Users

**Fig. 2.** Lorenz Curve for the Revenue

We use the real sales data of the platform during 2016 for certificate services. The platform offers three kinds of certificate: 1) the *free electronic certificates* that users can download the PDF when they successfully pass the MOOC; 2) the *paper certificates* with counter-forgery marks but no signature tracks from the instructors, and the unit price is 100RMB (about US$14.5), including shipping and handling. 3) the verified certificate with counter-forgery marks, signature tracks and the authentication to the users, and the unit price is 300RMB (about US$43).

There is a total of 1236 MOOCs available on the platform during 2016, and some MOOCs offer several times in different semesters, and we treat them different MOOCs. By the end of December, there are 1140 MOOCs already closed. We use the sales data of the closed courses in the analysis. Due to confidentiality reasons, we only show some basic statistical results.

### 4.1 Differentiation and Inequality of Revenue Generation

Figure 1 shows the relationship between the active users and the paying users. The *active users* are the users who have learning activities such as doing the homework, and the paying users are the users who buy either the paper certificate or the verified certificate. We can see that most courses are located at the bottom-left corner with a small number of active users and paying users. There is no direct relationship between the number of active users and paying users, and many factors of the courses such as difficulties, popularities, and practicability, may affect the relationship.

We also use the *Lorenz Curve* [14] to show the inequality of the revenue generation for different MOOCs. Figure 2 shows the Lorenz Curve for the revenue of the MOOCs on the platform. We can calculate that the *Gini coefficient* for the certificate market is 0.838, and the top 15% of the most profitable MOOCs create more than 80% of the total revenue for B2C services.

Based on our model in Section 3, we use the following method to infer the WTP of the users from the sales data: Recall the definition $V_j$ as the WTP of user $j$ for the certificate. If user $j$ complete the course, we assume $V_j > 0$; if user $j$ buys a paper certificate, then $100 \leq V_j < 300$; if user $j$ buys a verified certificate, then $V_j \geq 300$. From the sales data, we get the number of users with their WTPs located in the three intervals of $(0, 100)$, $[100, 300)$ and $[300, \infty)$. Now we show some insights of analyzing the data based on our model.

**Table 1.** WTP distributions for the best-selling MOOCs.

| Subject Category | Completion Rate | $WTP > 0$ | $100 \leq WTP < 300$ | $WTP \geq 300$ |
|---|---|---|---|---|
| Accounting | 2.9% | 870 | 315 | 381 |
| Marketing | 1.3% | 362 | 73 | 69 |
| Entrepreneurship | 1.2% | 385 | 48 | 63 |
| Accounting | 1.6% | 110 | 24 | 48 |

**Table 2.** WTP distributions of offering a MOOC multiple times.

| Semester | Completion Rate | $WTP > 0$ | $100 \leq WTP < 300$ | $WTP \geq 300$ |
|---|---|---|---|---|
| Fall 2015 | 2.9% | 870 | 315 | 381 |
| Spring 2016 | 1.3% | 566 | 184 | 236 |
| Summer 2016 | 1.8% | 257 | 73 | 99 |

### 4.2 The Best-Selling MOOCs

We observe the data from the best-selling courses and estimate the user's WTP. The top four best-selling MOOCs are all economic and management courses: two accounting courses, one marketing course, and one entrepreneurship course.

We find that the best-selling economic courses have similar WTP distributions. More users prefer the verified certificate to the paper certificate for each course, which shows that the paying users care more about the quality of service when the course is popular and useful.

### 4.3 Offer the Same MOOC Repeatedly

Our model in Section 3 indicates that the WTP distribution remains the same in different settings. We observe the case of offering a MOOC multiple times. We select the most popular accounting course on the platform and compare the sales data in three semesters.

We can see that the proportional relations of the three values for each semester are almost the same, which shows that the WTP distribution for a course does not change when we offer it multiple times, and indicating that the WTP distribution may be affected more by the intrinsic properties of the course (e.g. quality, usefulness) instead of the external factors. On the other hand, we also see that the total number of paying users declines when we rerun a course multiple times, indicating the *law of diminishing returns* [15] in economics.

Table 3. WTP distributions for the MOOCs with the highest payment rate.

| Subject Category | Completion Rate | $WTP > 0$ | $100 \leq WTP < 300$ | $WTP \geq 300$ |
|---|---|---|---|---|
| Financial Engineering | 0.24% | 21 | 3 | 16 |
| Computer Science | 0.45% | 42 | 12 | 69 |
| Mathematics | 0.82% | 9 | 3 | 5 |
| Computer Science | 0.35% | 29 | 8 | 17 |

### 4.4 MOOCs with the Highest Payment Rate

We then observe the sales data for the courses with highest *payment rate*, which is the proportion of paying users among the users completing the MOOC.

We observe that the MOOCs with highest payment rate are those science and engineering courses with high estimated efforts to complete. Also, the paying users for these courses have higher WTPs as they have already invested much time in the courses.

## 5 Future Work

In this section, we present our plans for improving our model for B2C markets by considering more factors affecting the MOOC profitability. We also propose the ideas of modeling the B2B course sub-licensing market.

### 5.1 Other Factors Affecting the B2C Markets

To make our model straightforward, we only adopt some basic variables to the models in Section 3. There are some other factors in consideration when we apply the models in industry and we will verify the relationship between these factors and the marketing performance in our future work.

**Growing User Bases:** The size of the user base of MOOC platform is dynamically increasing. In practice, we care more about the proportion of paying users among the active users than the actual number of paying users. Under this setting, the pricing strategy is a trade-off between the proportion of paying users and the net revenue gained from each user.

**Competitions among platforms:** The MOOC market is an oligopoly market with a limited number of competitive MOOC platforms, and each platform has millions of registered users, as small players lack the source of MOOC producers and can not afford the high production cost. There are always similar courses among different MOOC platforms (e.g. various kinds of data science MOOCs). The differentiation strategy is a fundamental way to attract the users. In addition, some platforms occasionally offer discounts for their value-added services, some platforms reorganize the courses into bundled courses, and some platforms offer more attractive value-added services.

**Externalities:** For MOOC services, We may also consider the case that users impose externalities on each other. For instance, there may be a positive externality in which a users learning outcome improves as other users give her more help on the discussion forum. On the other hand, one could also observe negative externalities, for instance, one user's resource consumption may affect the other users' experience of watching MOOC videos.

**Seasonality:** MOOC services have a strong seasonality. In winter and summer vacations, there are almost no B2B sub-licensing services to offer since the institutional users are on vacation. On holidays, individual users have more time on the Internet and the motivated learners will spend more time on taking MOOCs. At the end of each semester, MOOCs may conflict with the in-classroom courses of the college users, and thus attract less active users.

### 5.2 Modeling the B2B Market

Although the B2B model arises less attention to both the industries and academics in the MOOC ecosystem, it also plays an important role for the MOOCs, and brings more revenue to some early-stage MOOC platforms than B2C services. In the B2B *course sub-licensing* market, the MOOC platform is the seller in the market, and users are organizations with the demand of using the MOOC content on the platform for education purpose. Since the copyright of the licensed MOOCs does not belong to the platform, the MOOC platform should first get sub-licensing approval from the MOOC producers and share revenue with them.

The sub-licensing service is an excellent way to help universities improving their teaching outcomes by importing high-quality MOOC content from MOOC platform. In practice, the B2B services always exist in the format of *SPOCs* (i.e. Small Private Online Courses) by using blended teaching and learning approaches. To guarantee the quality of service, the sub-licensing services are always dynamic with highly customized bundles including various education services, such as MOOC contents, teaching assistance services, SaaS services, technical supports and so forth. As the users' demands are dynamic and complicated, we can no longer use the flat-rate pricing for the B2B markets. The auction-based pricing scheme can better fit the market settings. We will formulate and analyze the auction-based pricing scheme for the B2B market in our future work.

Another cutting-edge B2B business model is *cross-platform MOOC exchange and internationalization*, which is a hybrid B2B2C (i.e. business-to-business-to-customer) business model. The two platforms both benefit from the content and revenue sharing collaboration, as it will help each platform make money from some hard-to-reach secondary markets. An experiment is that edX sub-license some of its courses to xuetangX, by allowing xuetangX to translate the materials into Chinese and provide localized teaching assistant services [16].

## 6    Conclusion Remarks

Working in the MOOC industry for the past three years, we realize the importance of designing appropriate marketing strategies to sustain our business. Even though there is few academic work on analyzing the business models for MOOCs in scientific and structured ways, we make our initial attempts to construct mathematical models to capture the insights of MOOC markets and use data-driven marketing approaches to verify our models. We hope that our current and future research will bring more ideas to both the industries and the academics for the sustainable development of MOOC ecosystems.